# Disjoining Pressure of Water in Nanochannels

*An Zou, Sajag Poudel, Manish Gupta and Shalabh C. Maroo*[*]
Department of Mechanical and Aerospace Engineering, Syracuse University, Syracuse, NY 13244
[*]corresponding author: scmaroo@syr.edu

**Abstract:** Disjoining pressure of water was estimated from wicking experiments in 1D silicon-dioxide nanochannels of heights 59 nm, 87 nm, 124 nm and 1015 nm. The disjoining pressure was found to be as high as ~1.5 MPa while exponentially decreasing with increasing channel height. Such a relation resulting from curve fitting of experimentally-derived data was implemented and validated in computational fluid dynamics. The implementation was then used to simulate bubble nucleation in water-filled 59 nm height nanochannel to determine the nucleation temperature. Simultaneously, experiments were conducted by nucleating a bubble in a similar 58 nm height nanochannel by laser heating. The measured nucleation temperature was found to be in excellent agreement with the simulation, thus independently validating the disjoining pressure relation developed in this work. The methodology implemented here integrates experimental nanoscale physics into continuum simulations thus enabling numerical study of various phenomena where disjoining pressure plays an important role.

**Keywords:** Nanochannel, Water, Disjoining pressure, Wicking, Nucleation, Capillary Pressure

**Main Text**

A nanoscale thin liquid film on a surface can have significantly different properties than its bulk form.[1] At such short distances, intermolecular interactions with surface atoms can dominate and define new equilibrium positions/velocities of liquid atoms; as these fundamental parameters are statistically averaged to estimate thermodynamic properties,[2] substantial changes in density, pressure, surface tension, viscosity, etc. can occur. Distances upto which a surface can affect liquid properties depends on the atomic composition: if either atom is non-polar, the presence of only weak and short-range van der Waal's force limits such changes to <5 nm;[3, 4] however, if both atoms are polar, strong and long-range electrostatic forces can alter properties up to tens to hundreds of nanometers from the surface.[5-7] The latter scenario often occurs in practical situations involving water on various surfaces.

In this work, we focus on the pressure of nanoscale water films. The thermodynamic equilibrium of liquid/air inside the nanochannels is dependent on several parameters including van del Waals forces, electrostatic forces, and structural forces, thus it represents a Gibbsian composite system.[8] Here, we use disjoining pressure as the major variable governing the alteration in nanoscale liquid film, which also lays the foundation for advancing research on liquid-vapor phase-change heat transfer to develop future thermal management devices. The pressure in such thin films is expected to be reduced based on the modified Young-Laplace equation.[9] Such a reduction, defined as disjoining pressure,[10] plays a fundamental role in a wide range of engineering and natural systems involving bubbles,[9, 11-15] transpiration,[16, 17] emulsions,[18-22] and membranes.[23] Theoretical determination of disjoining pressure of water on surfaces using extended DLVO theory requires fitted constants as the structural forces are usually unknown, and the surface potential cannot be measured directly for deionized (DI) water.[21, 24-29] On the other hand, numerical simulations such as molecular dynamics are currently inept to computationally simulate large domains and accurately capture the intermolecular forces of water over larger distances. Likewise, the experimental estimation of sub-100 nm liquid films has been severely limited to non-polar films on solid surfaces[30, 31] primarily due to the evaporative and fluidic nature of nanoscale water films. Here, we overcome these experimental challenges by characterizing water's wicking behavior in 1D silicon-dioxide ($SiO_2$) nanochannels of varying height, estimating the disjoining pressure of water from experimental data, and applying these disjoining pressure values in computational fluid dynamics (CFD) wicking and bubble nucleation simulations.

Figure 1a shows the sketch of a typical sample used in our experiments. The nanochannels are ~2 cm long and 10 μm wide, with a 10 μm spacing between two adjacent channels (please refer to supporting information for sample fabrication). Due to the long-range ordering (up to 50 nm away from surface) of water molecules on silicon-dioxide surface shown in nuclear magnetic resonance studies,[6, 7] we fabricated four channel heights of 59, 87, 124, and 1015 nm (Fig. 1b). A ~50 μm deep reservoir was etched at each end of the channels. To conduct experiments, a DI water drop was placed in one reservoir causing the water to wick into the nanochannels and reach the other end. The top view of this wicking process was recorded using a high speed camera, where a wicking distance $l$ was defined as the distance between the entrance and the position of the meniscus at a given time $t$ (Fig. 1c). A microscope with a 50× objective was optionally connected to the camera to obtain a close-up view of the meniscus during wicking. After the experiment, the sample was dried by heating it to ~200 ºC on a hot plate in open air with the drying time ranging from several minutes up to several hours depending on the nanochannel height. Wicking experiments for each channel height were conducted at least four times to ensure repeatability.

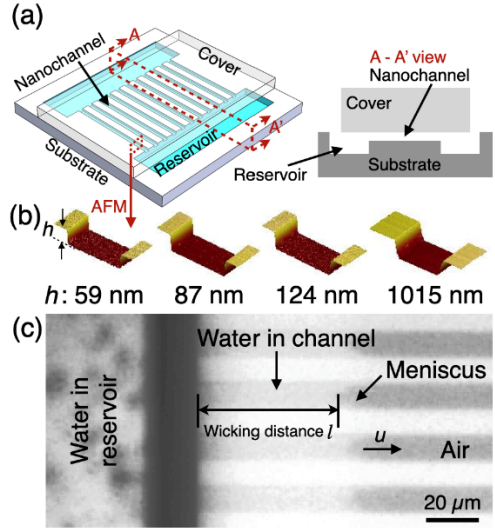

**Figure 1:** (a) Sketch of the sample with nanochannels and reservoirs for wicking experiments. (b) Atomic force microscope (AFM) images of the fabricated nanochannels of different heights. The AFM images were taken right before the bonding process which closes the top of the channels. (c) Optical microscope image of water wicking into nanochannels of height 59 nm; the black dots in reservoir are roughness introduced from deep silicon etching during sample fabrication while the black bar at the entrance is the edge of the reservoir which is out of focus.

The wicking distance $l$ variation with time $t$ is plotted in Fig. 2; $l$ is found to be proportional to $t^{1/2}$ for all channel heights as predicted by the analytical solution of Navier-Stokes equation for a high aspect ratio (height << width) rectangular cross-section channel:[32]

$$l = \sqrt{\frac{h^2\left[1 - 0.63\left(\frac{h}{w}\right)\right]\Sigma P}{6\mu}} \cdot t^{1/2} \approx \sqrt{\frac{h^2 \Sigma P}{6\mu}} \cdot t^{1/2} = C \cdot t^{1/2} \quad (1)$$

where $h$ and $w$ are height and width of the nanochannel, respectively; $\mu$ is the fluid dynamic viscosity; and $\Sigma P$ is the total pressure difference driving the wicking. The value of constant $C$ (i.e. the slope in Fig. 2) is obtained from the linear curve fitting of experimental data in Fig. 2 for the highest $R^2$ value, and $C$ increases with increasing channel height as expected. During the wicking process, two types of menisci were observed: regular-curved shape and wedge shape (insets in Fig. 2). In 1015 nm channel height, the meniscus was always of regular-curved shape as it moved in the entire channel length (~2 cm). However, in 59 nm and 87 nm height channels, the regular-curved meniscus occurred only within the first few hundreds of microns from the entrance. Beyond this initial distance, water was seen to flow faster at the corners forming the wedge-shape meniscus, which also led to air or vapor being momentarily trapped within the wicking liquid front.[33] Meanwhile, the 124 nm height channel was transitional as both types of menisci simultaneously co-existed in neighboring channels even at lengths far away from entrance (~1.5 cm).

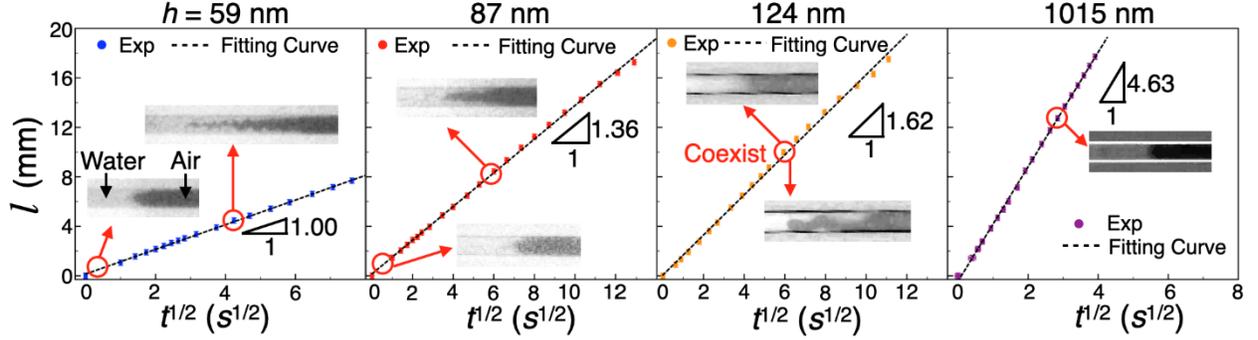

**Figure 2:** Variation of wicking distance $l$ with $t^{1/2}$ in 59 nm, 87 nm, 124 nm, and 1015 nm channels. The insets are top-view images from optical microscope showing the two different meniscus shapes observed in the experiments.

The most common mechanism of wicking in nanochannels is attributed to capillary pressure.[34-37] If wicking is only driven by capillary pressure (i.e. $\Sigma P = P_c \approx 2\sigma cos\theta/h$) where $\sigma$ is the surface tension, and $\theta$ is the contact angle, Eq. 1 can be simplified to the widely-used Washburn equation for capillary filling:[38]

$$l = \sqrt{\frac{\sigma h cos\theta}{3\mu}} \cdot t^{1/2} \quad (2)$$

Although the linear dependence of wicking distance on $t^{1/2}$ holds at nanoscale, Washburn equation is inconsistent in predicting the experimental wicking rate. Other published literature on wicking in rectangular cross-section nanochannels have found a similar observation[33, 39-42] and explained this inconsistency primarily due to electro-viscous effect[42-45] or geometrical effect.[46, 47] However, these effects do not explain our observed deviation (please refer to supporting information for detailed explanation). Further, a common major flaw in these prior studies is the contact angle $\theta$ used in Eq. 2. We should not inherently assume a uniform contact angle[33, 42, 48-50] of the meniscus due to the channel's high aspect ratio (width is ~10 μm for all channels while height varies from 0.059 μm to 1.015 μm). Thus, due to the rectangular cross-section of the channel, we define two separate contact angles $\theta_{top}$ and $\theta_{side}$ of the meniscus corresponding to the top and side views, respectively (Fig. 3a).

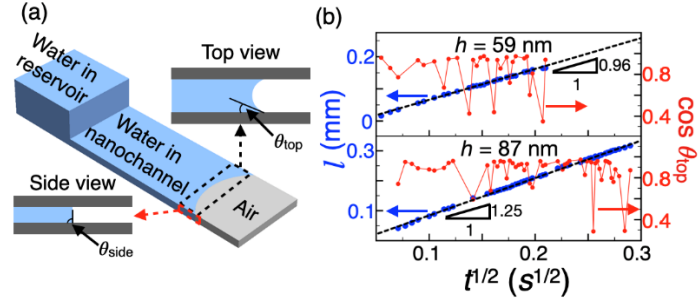

**Figure 3:** (a) Sketch showing the difference in contact angles from top and side views. (b) Measured wicking distance and $cos\,\theta_{top}$ as a function of $t^{1/2}$ in 59 nm and 87 nm channels during the initial time when the meniscus was only regular-curve shaped.

Figure 3b plots the change in $\theta_{top}$ for the 59 nm and 87 nm channel heights along with their respective wicking rates during the initial time period when the meniscus is regular-curved shape (please see supporting information for $\theta_{top}$ estimation from optical microscope images). The experimental data presents two unique findings: (1) despite as much as 50% fluctuation in the values of $cos\,\theta_{top}$, the wicking distance $l$ is still proportional to $t^{1/2}$ (Fig. 3b), and (2) the fitting curve slopes of 0.96 and 1.25 mm/s$^{1/2}$ for 59 nm and 87 nm channels, respectively, during the initial wicking period (Fig. 3b) are in good agreement with the values for the entire wicking period (1.00 and 1.36 mm/s$^{1/2}$ respectively in Fig. 2) even though the meniscus shape drastically changes from regular-curved shape to wedge-shape. These observations imply that capillary pressure, which is strongly dependent on meniscus shape, does not govern the wicking phenomenon in smaller channel heights of 59 nm and 87 nm. Instead, disjoining pressure must be the driving force as most or all of the water in these nanochannels interact with the solid surface due to the small height. The opposite holds true for the larger channel height of 1015 nm where only the steady regular-curved shape meniscus occurs implying that capillary pressure drives wicking, as merely a small fraction of the total water will be affected by the surface. In the transitional 124 nm height channel, both disjoining and capillary pressures are important. The importance of both pressures at this channel height also alludes to the idea that the effect of $SiO_2$ surface on water (i.e. disjoining pressure) starts to diminish at distances beyond $h/2\sim60$ nm from the surface, and is in good agreement with published studies[6, 7] which found the distance to be ~50 nm using nuclear magnetic resonance. Next, we use the experimental data to calculate disjoining pressure in the nanochannels.

Disjoining pressure is theoretically known to continuously decrease with increasing distance from the surface.[51] However, as we only have four channel heights, disjoining pressure $\overline{P}_d$ averaged over half the channel height is estimated from experimental data using Eq. 1 and $\Sigma P = \overline{P}_d + P_c$, the combination of which can be rearranged into:

$$\overline{P}_d = \Sigma P - P_c = \frac{6\mu C^2}{h^2} - 2\sigma\left(\frac{cos\theta_{side}}{h} + \frac{cos\theta_{top}}{w}\right) \quad (3)$$

where $C$ is the slope (from Fig. 2) of the wicking rate and $\theta_{top}$ is obtained from top view images as mentioned earlier. It should be noted here that although properties of water ($\mu$ and $\sigma$) in such confined channels will be different from the bulk, we still assume bulk properties as per the norm in literature[40-43, 46] and also due to the fact that such properties have not yet been directly measured in experiments; in fact, disjoining pressure term is introduced in the modified Young-Laplace equation[15, 30] to account for the lack of such properties at these scales. Thus, in Eq. 3, all parameters are known except $\theta_{side}$. However, as the heights of the channels are at nanoscale, it is not possible to experimentally measure $\theta_{side}$. Considering the published experimental finding that water molecules on a glass surface are highly ordered upto distances of 50 nm normal to the surface[6, 7] (i.e. channel heights of upto 100 nm in our case), we use molecular dynamics simulations to qualitatively study the effect of liquid structuring on contact angle by simulating wicking of liquid argon in a hydrophilic nanochannel of two different heights (please see supporting information for simulation details and results). For the smaller channel height where liquid structuring is prominent, $\theta_{side}$ is obtained to be ~90°, whereas the larger channel height with mostly bulk liquid resulted in $\theta_{side}$ of ~45°. Thus, using the same analogy for water in our SiO$_2$ nanochannels, we assume $\theta_{side}$ as 90° for the smaller height channels (59 nm and 87 nm) and $\theta_{side} = \theta_{top} = 39.6°$ for larger channel height (1015 nm) in Eq. 3 to estimate $\overline{P}_d$ (Fig. 4a). For the transitional channel with the height of 124 nm, $\theta_{side}$ would be between 90° and $\theta_{top}$; hence, we take the average of upper and lower $\overline{P}_d$ values based on this range of $\theta_{side}$ while simultaneously showing the limits (Fig. 4a). Please see supporting information for the values of $\overline{P}_d$ and $P_c$ for each channel height. Thus, $\overline{P}_d$ of water in silicon-dioxide nanochannels is found to exponentially decrease with increasing water film thickness δ (Fig. 4a) which is equivalent to half the channel height $h/2$. We choose the exponential relation as polar molecules exhibit dominant electrostatic interaction as compared to power law function when only van der Waals interaction is present.[21] Such a relation can be expressed as:

$$\overline{P}_d = A \cdot e^{-b\delta} \quad (4)$$

where A = (4.25 ± 1.06)×10$^6$ Pa, and $b$ = 0.035 ± 0.007 nm$^{-1}$ with 50% confidence. The values of constants A and b are obtained from curve fitting the experimental data of Fig. 4a, where the gray region shows the range of 50% confidence bounds.

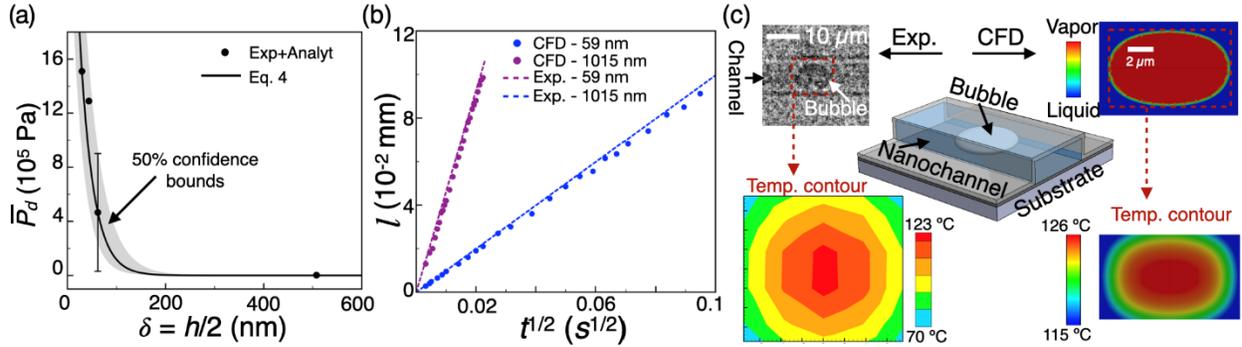

**Figure 4:** (a) Estimated average disjoining pressure $\overline{P}_d$ of water in silicon-dioxide nanochannels for various water film thickness δ based on nanochannel height h, and the corresponding exponentially fitted curve with R$^2$ = 0.97. The gray region represents the 50% confidence bounds. (b) Comparison of experimental wicking rate with that from CFD simulations where average disjoining pressure data of (a) is integrated into the simulations. (c) Sketch (center) of bubble nucleation in 59 nm nanochannel with experimental image (top left) of the bubble and corresponding measured temperature distribution (bottom left); phase contour plots from CFD simulation of bubble nucleation in nanochannel (top right) with the corresponding temperature distribution (bottom right). The colored bubble image was converted to gray scale and its contrast was enhanced using MATLAB functions.

Next, we implement Eq. 4 in commercial CFD software ANSYS Fluent[52] to show the applicability of experimentally derived $\overline{P}_d$ values into continuum simulations. Similar to the experiments, we simulate wicking of water in 59 nm and 1015 nm channels (please see supporting information for details) using a laminar multiphase volume of fluids (VOF) method.[53] As the driving force of the wicking is $\Sigma P = \overline{P}_d + P_c$, $P_c$ is achieved through continuous surface force (CSF) modeling[54] along with wall adhesion and $\overline{P}_d$ is invoked by a user-defined-function (udf). The different $\theta_{top}$ and $\theta_{side}$ in 59 nm channel are implemented by assigning the corresponding contact angles to side walls and top/bottom walls of the channel, respectively. Wicking rates are found to be in good agreement with experimental data in both channels (Fig. 4b) demonstrating successful implementation of disjoining pressure model in CFD. Such an integration captures nanoscale physics in continuum simulations while including the effects of disjoining pressure in various phenomena such as phase change, transpiration, etc.

In order to quantitatively and independently validate the disjoining pressure expression (Eq. 4), we performed experiments and continuum simulations of bubble nucleation inside a silicon-dioxide nanochannel and compared the nucleation temperature values. For the simulations, we first developed an expression[55] for local disjoining pressure $P_d$ as a function of the distance from surface, $P_d = 5.765 e^{-0.142 x}$, from the average disjoining pressure equation (Eq. 4) and implemented the expression in ANSYS Fluent as a user defined function (similar to the implementation mentioned above which resulted in Fig. 4b). We simulated bubble nucleation in water confined in a 59 nm height nanochannel by applying a localized heating source with constant heat flux (Fig. 4c). Initially, the nanochannel domain was completely filled with water at 300 K. A constant heat flux was then supplied at the specified spot of the bottom surface which increased the liquid temperature inside the domain (please refer to supporting information for details of CFD simulation). The bubble nucleation temperature was obtained as 126.0 ºC in the CFD simulation (Fig. 4c). Due to the disjoining pressure effect, the liquid pressure in such nanochannel is higher[17, 56] than bulk liquid pressure, thus requiring higher temperature to nucleate a bubble in water-filled-nanochannel than bulk water. To validate the CFD simulation result, nanochannels of 58 nm height were fabricated with multiple layers buried underneath the nanochannel; the layers assist in absorbing an incident laser beam and heating the surface to nucleate a bubble.[11, 12] A typical bubble nucleation process in a 103 nm height nanochannel is shown in supporting information video as it provides a much better contrast and visualization than nucleation in 58 nm height nanochannel (for which MATLAB processing was required as shown in Fig. 4c). The nucleation temperature in nanochannel was measured directly using infrared (IR) camera with a 4× objective and was found to be 123.9 ± 3.0 ºC (Fig. 4c, please see supporting information for details of experiments), in excellent agreement with CFD simulations. Thus, the obtained disjoining pressure expression (Eq. 4) is independently validated, and can be further utilized to explore transport phenomena in various nanofluidic applications.

To summarize, we report a fundamental study of estimating disjoining pressure of water in silicon-dioxide nanochannels through experiments of wicking in channels of heights 59 nm, 87 nm, 124 nm and 1015 nm. Disjoining pressure is found to be the primary driving force of wicking in smaller height nanochannels while capillary pressure dictates wicking in larger height nanochannel. The average disjoining pressure of water exponentially decreased with increasing distance from the surface, and a relation is derived by curve fitting the experimental data. The disjoining pressure relation is implemented in CFD simulations and is shown to capture the experimental wicking behavior. Such an implementation is then used to simulate bubble nucleation in water filled nanochannel of height 59 nm. Simultaneously, nanochannels of height 58 nm were fabricated and bubble nucleation was achieved inside the water-filled-nanochannel by laser heating. The bubble nucleation temperature measured in experiments was found to be in excellent agreement with that obtained from CFD simulations, thus independently verifying the disjoining pressure model developed in our work.

## Supporting Information

Additional details on sample fabrication, wicking rate in nanochannels, dynamic contact angle, MD simulation, disjoining pressure calculation, CFD simulation of wicking, and nucleation temperature of water in nanochannel.

Experimental video of bubble nucleation inside nanochannel.

## Notes

The authors declare no competing financial interest.

**Acknowledgements**

This material is based upon work supported by, or in part by, the Office of Naval Research under contract/grant no. N000141812357. This work was performed in part at the Cornell NanoScale Facility, an NNCI member supported by NSF Grant NNCI-2025233.

**Table of Contents Graphic**

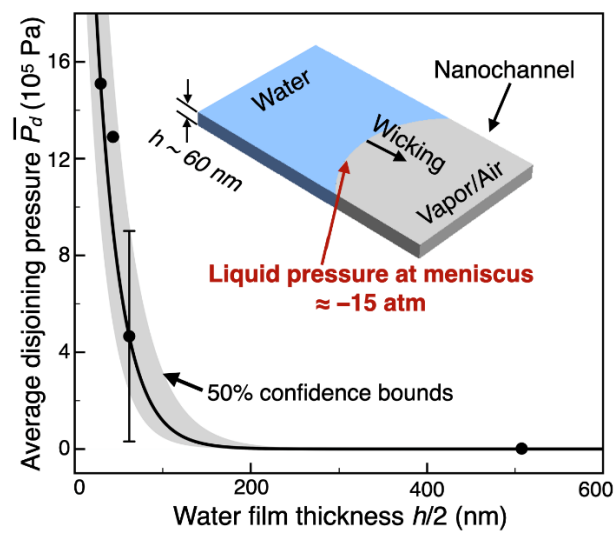